**Dynamic parameters of structures extracted from ambient vibration measurements: an aid for the seismic vulnerability assessment of existing buildings in moderate seismic hazard regions**

Clotaire Michel[1], Philippe Guéguen[1,2] and Pierre-Yves Bard[1,2]

[1] Laboratoire de Géophysique Interne et Tectonophysique, University of Grenoble, France

[2] Laboratoire Central des Ponts-et-Chaussées, Paris, France

Corresponding author, cmichel@obs.ujf-grenoble.fr
    Tel: +33 4 76 82 80 71
    Fax: +33 4 76 82 81 01

**Abstract**

During the past two decades, the use of ambient vibrations for modal analysis of structures has increased as compared to the traditional techniques (forced vibrations). The Frequency Domain Decomposition method is nowadays widely used in modal analysis because of its accuracy and simplicity. In this paper, we first present the physical meaning of the FDD method to estimate the modal parameters. We discuss then the process used for the evaluation of the building stiffness deduced from the modal shapes. The models considered here are 1D lumped-mass beams and especially the shear beam. The analytical solution of the equations of motion makes it possible to simulate the motion due to a weak to moderate earthquake and then the inter-storey drift knowing only the modal parameters (modal model). This process is finally applied to a 9-storey reinforced concrete (RC) dwelling in Grenoble (France). We successfully compared the building motion for an artificial ground motion deduced from the model estimated using ambient vibrations and recorded in the building. The stiffness of each storey and the inter-storey drift were also calculated.

# 1. Introduction

Knowing the dynamic parameters of a structure (e.g. a building or a bridge) may be useful: (1) to calibrate its elastic properties for numerical modelling, (2) to detect the modification of its behaviour after retrofitting or damage, (3) and, finally, to predict its behaviour under earthquakes. The linear dynamic behaviour can be fully described by the modal parameters: resonance frequencies, modal shapes and damping ratios. These parameters mainly depend on the storey masses, which remain unchanged whatever the state of the structure, and the storey stiffnesses, which are influenced by the structural modifications such as reinforcing and damage. It can also be representative of the quality of the material (e.g. the equivalent Young's modulus of the cracked or undamaged concrete) and of the structural design (e.g. irregularity of the shear resistance or soft storey). All large-scale vulnerability assessment methods (e.g. [1, 2, 3]) identify the stiffness regularity of buildings as one of the main parameters controlling their seismic resistance. The stiffness is also the basic parameter needed to draw the capacity curve of buildings, which is the basis of recent large-scale vulnerability assessment methods (e.g. [1, 3, 4]). One of the greatest difficulties within such methods is the lack of information on existing buildings. Within vulnerability assessment, we have to deal with questions on, for example, ageing, structural design, quality of materials and the building state.

Assessing the stiffness of each storey, and therefore the modal shapes of the structure, is then a critical point for the evaluation of the dynamic properties of existing buildings and hence to predict their seismic response and vulnerability. The ratio between transverse and longitudinal frequencies can, for example, provide information on the relative stiffness in perpendicular directions that may expose a lack of stiffness and may help in structural design understanding.

Comparing the frequencies and mode shape before and after the shaking [5, 6, 7] allows the evaluation of damage. This method requires knowledge of the initial state of the structure in such a way that the modification due to damage can be observed. For example, comparisons of frequencies among a group of identical buildings can quantify the damage level of each building [8]. Based on the integrity threshold concept [9], and with only limited structural information, the integrity of the building under seismic shaking can be estimated using the experimental modal shape deduced from forced or ambient vibrations.

Ambient vibrations provide information about the modal parameters of a structure that we can extract using Modal Analysis methods. There are many different techniques to identify the "natural" modes of a structure, which can be divided into parametric and non-parametric methods. In the first category, the parameters of the considered model are updated to fit the recorded data in frequency or in time domain (see [10] for details). In the second category, the methods use only signal processing tools so that they are more user-friendly and easier to implement. Moreover, existing civil engineering structures are often difficult to model relevantly so that it is necessary to test these approaches before any more complicated method. For example, the Peak Picking (PP) method consists in taking the frequency peaks of average spectra for each sensors placed at different points. The Frequency Domain Decomposition (FDD [11]) is an improvement of the PP method. It consists of decomposing the power spectral density matrices into single-degrees-of-freedom systems by singular value decomposition.

The main goal of this paper is to demonstrate the utility of techniques using ambient vibration recordings to complement and improve seismic vulnerability investigations of existing buildings. Ambient vibrations recorded in buildings are processed for estimating the dynamic parameters of the building based on modal parameter estimates (Fig. 1). Because of its low cost and the efficiency of operational modal analysis, the ambient vibration method is well adapted to large-scale analysis for which a large set of buildings has to be analyzed. We first describe the FDD method we used in this study, one method among a large set of others existing methods. Then we propose the shear beam model, adapted to our recording layout to approximate the motion of a building. An analytical solution is proposed to estimate the stiffness at each story as a function of the modal shapes. We apply this theoretical work to a building located in Grenoble (France). We estimate its modal properties using ambient vibration tests and recordings of a ground motion induced by a bridge demolition in the same area. We then validate the modal model derived using modal parameters estimated from ambient vibrations by comparing the simulation results of the bridge collapse with real data. Finally we estimate the stiffness at each story of this building and the inter-story drift during the seismic motion and deduce some information about its seismic behaviour.

**2. Modal parameters from ambient vibration recordings**

In the 1930s, US Coast and Geodetic Survey first undertook ambient vibration recordings in high-rise buildings and bridges to determine their fundamental periods [12, 13]. Until the late 1990s engineers preferred forced vibration tests (Traditional Modal Analysis) because of the accuracy of the corresponding system identification techniques. However, during the last two decades ambient vibration recordings have become to be preferred because of their low cost. They are used to determine the behaviour of structures (frequencies and modal shapes) [14, 15], to quantify the damage after earthquakes [16, 17, 8] and to assess the benefit of retrofitting [18, 19]. They also allow the calibration of numerical and analytical modelling (model updating) [15, 20, 21]. Ambient vibrations are produced by natural sources such as local atmospheric conditions (e.g. the wind and the sea) or by human activities (e.g. traffic and factories). The expected range of acceleration values for ambient vibration tests is $10^{-7}$ to $10^{-4}$ g. Therefore, there is no doubt that the low level of the shaking only gives relevant information on the elastic behaviour of the structure. Many authors [e.g., 17, 18, 19, 20, 23, 24, 25, 26] showed that non-linearity affects resonance frequencies when the level of loading increases. This non-linearity is due to the opening of micro-cracks in the concrete during vibrations, which may decrease temporarily, or permanently the stiffness, and then the resonance frequencies, of the structure. Nevertheless, recent studies showed this decrease may be low for weak to moderate motions [17] and is usually less than 20%, which confirms the interest of conducting ambient vibration recordings within buildings.

Recording ambient vibrations at different points of a civil engineering structure (e.g. a bridge, a building or a chimney) allows the determination of its modes of vibration through Operational Modal Analysis techniques [27]. The efficiency of the Output-only Modal Analysis algorithms, the low cost of ambient vibration tests and the wide range of potential applications are the main reasons why their use is widespread today.

Practical issues for vibration recordings are, on one hand, the layout of measurement arrays and, on the other hand, the frequency and time length of recordings. The number of recorded points depends on the height, complexity and accessibility of the building. For example, in dwellings, only the stairs are usually available so that motions can only be recorded at one point per floor. The drawback is that it is impossible to discriminate torsion modes in that case. The frequency of sampling depends on the greatest frequency we want to estimate, practically never more than 30 Hz for usual buildings. Moreover, Brincker [27] showed that in practice the time length of the recorded window should be at least equivalent to 1000 periods of the structure in order to calculate an accurate spectral estimate.

Contrary to forced vibrations and earthquakes, the excitation is transmitted by external forces spread along the structure (wind), by the ground (seismic ambient noise) and by internal forces (human occupancy). Hence, it is impossible to determine the input load. The techniques used to determine the modal parameters of the considered system are called Output-Only or Operational Modal Analysis [10]. Many different techniques exist, the most widely used and easiest one being the Peak Picking method (PP). It consists of calculating the Fourier transforms of short time windows (several seconds) and picking the value of the frequency peaks of the average spectrum. The amplitude of the peak for each simultaneous recording gives a point of the modal shape. The normalised shape is obtained by dividing it by the value at the reference sensor. Theoretically, it only gives the operational deflection shapes, which are not characteristics of the system but depends on the input. Practically, PP can give accurate estimates of the modal shapes if the modes are well separated, that is often the case for regular tall buildings.

Many other methods in time or frequency domains have been developed under the theory of System Analysis. The easiest one is called Frequency Domain Decomposition (FDD) [11] and consists of decomposing the power spectral density matrices into single-degrees-of-freedom systems by singular value decomposition. The initial description of the FDD method can be found in [11]. The Power Spectral Density (PSD) matrices are the Fourier transforms of the correlation matrices between all the simultaneously recorded signals so that no *a priori* model is supposed. The PSD matrices of the input (unknown) and the output (recorded) signals, functions of the angular frequency ω, can be respectively noted $[X](\omega)$ and $[Y](\omega)$. They are linked to the frequency response function matrix $[H](\omega)$ through the equation:

$$[Y](\omega) = [\overline{H}](\omega)[X](\omega)[H](\omega)^T \tag{1}$$

where $^T$ denotes transpose and $\overline{H}$ is complex conjugate. If $r$ is the number of inputs and $m$ the number of simultaneous recordings, at each angular frequency ω, the sizes of [X], [Y] and [H] are $r \times r$, $m \times m$ and $m \times r$, respectively. In Operational Modal Analysis, the usual assumption is that the input is white noise, which means:

$$[X](\omega) = C \tag{2}$$

which is a constant matrix. The [H] matrix can be written in a pole ($\lambda_k$) / residue ($[R_k]$) form as:

$$[H](\omega) = \sum_{k=1}^{n} \frac{[R_k]}{j\omega - \lambda_k} + \frac{[\overline{R_k}]}{j\omega - \overline{\lambda_k}} \tag{3}$$

where $j^2 = -1$.

Using Eq. 2 and 3 and for low values of damping, it can be shown that the term $[R_k][C][\overline{R_k}]$ dominates the expression of [Y] (Eq. 1). In this case, this term becomes also proportional to the mode shape vector $\{\phi_k\}$:

$$[R_k][C][\overline{R_k}] = d_k \{\Phi_k\}\{\Phi_k\}^T \tag{4}$$

where $d_k$ is a constant. Only a limited number of modes (typically no more than two) noted Sub(ω) has energy at one particular angular frequency ω. The PSD matrices of the outputs (Eq. 1) then has the following form:

$$[Y](\omega) = \sum_{k \in Sub(\omega)} \frac{d_k \{\Phi_k\}\{\Phi_k\}^T}{j\omega - \lambda_k} + \frac{\overline{d_k}\{\overline{\Phi_k}\}\{\overline{\Phi_k}\}^T}{j\omega - \overline{\lambda_k}} \tag{5}$$

In practice (see Fig. 2 for an illustrated case for synthetic data), the ambient vibrations recordings allow the estimation of the PSD matrices of the outputs $[\hat{Y}](\omega_i)$ at each known frequency $\omega_i$, i.e. the Fourier Transforms of the cross-correlation matrices. We used Welch's method [28] for this estimation, which means we took the average square spectrum of moving Hamming windows in the signal. The length of the windows is directly linked to the precision in the frequency we require. The common number of points in the Fourier Transform we used was 8192, which means windows of 20 s for a 200 Hz sampling frequency, corresponding to a frequency step of 0.024 Hz.

The PSD matrices cannot usually be diagonalised because at one frequency, only few (*p*) modes have energy so that the rank of the matrix is *p*, the other "eigenvalues" are close to zero (noise). That is why we have to perform a singular value decomposition of these matrices, such as:

$$[\hat{Y}](\omega_i) = [U_i][S_i][\overline{U_i}]^T \qquad (6)$$

where $[U_i]$ is the matrix of the singular vectors and $[S_i]$ is the diagonal matrix of the singular values. Close to a peak, if only one mode is dominating so that there is only one term in Eq. 5, the first singular vector is an estimate of the modal shape. If two modes are dominating at this frequency peak, which means two modes are close in frequency, and if they are geometrically orthogonal, the estimates of the corresponding modal shapes are the two first singular vectors. Moreover, the estimate of the two close mode shapes is still unbiased. In conclusion, the FDD method allows the accurate extraction of the mode shapes from ambient vibrations recorded simultaneously at several points of the structure. Several sets of recordings can give the mode shapes of the whole structure since a reference sensor allows the normalisation of the shape.

Many peaks may appear in the first singular value computed from ambient vibrations, not only those corresponding to structural modes. The Modal Assurance Criterion (MAC) [29] gives a quantitative value to compare two modal shapes $\Phi_1$ and $\Phi_2$ through the following expression:

$$MAC(\Phi_1,\Phi_2) = \frac{\left|\Phi_1^H \Phi_2\right|^2}{\left|\Phi_1^H \Phi_1\right|\left|\Phi_2^H \Phi_2\right|} \tag{7}$$

where $^H$ denotes complex conjugate and transpose. The MAC has been used in order to evaluate the damage by comparing the initial and the final mode shapes [5, 6] as well as to compare modes obtained by different ways (e.g. different methods or recordings). Moreover, comparing the modal shape at a given frequency peak with the modal shape at the surrounding frequencies can give the extent of the "bell" representing the corresponding single degree of freedom (Fig. 3). In practice, as long as the MAC value between the shape at one frequency and the shape at the peak is greater than 80%, we consider that the frequency belongs to the mode. By this way, it is possible to determine the frequency bandwidth where each mode is dominating and to discriminate peaks that do not correspond to a structural mode.

## 3. Stiffness deduced from modal parameters

### 3.1 Modelling of the system

As the mass per story can be considered as constant, the main parameter controlling the mode shape is the stiffness. The mode shapes of the system (building) are `unscaled' parameters, which are linked to the `scaled' physical parameters used in earthquake engineering (e.g. forces, acceleration, Young's modulus and inertia). For that reason, in order to derive the physical parameters of the structure from the modal parameters, a model is necessary. Based on the knowledge of the structural parameters of the building, the modal parameters and the output parameters (i.e. stress and strain in the building) are deduced from forward modelling.

On the other hand, the *a priori* structural parameters of the model have to be updated by solving the inverse problem [21], based on the knowledge of the modal parameters.

Modelling techniques using the Finite Element Method (e.g. [15, 30]) and homogenization [9] can be found in literature. The updating procedure can use only the first frequency to check if the others are well represented by the model or all the measured modal parameters. In this paper, we use simple elastic lumped-mass modelling because of its low-cost for a large set of buildings. Moreover, it does not require detailed information on the building (e.g. the plan, structural design and quality of materials), which is usually not available for existing buildings. Since the modal parameters are deduced using, at most, a few ambient vibration recordings per floor, the elastic lumped-mass modelling is accurate enough. It amounts to neglecting the internal deformation of the floors and in considering the deformation between two stories as linear.

Considering displacement in only one direction, Newton's second law for the displacement of the considered degrees of freedom (DOF) $\{V(t)\}$ can be written as follows:

$$[M]\{V''(t)\} + [C]\{V'(t)\} + [K]\{V(t)\} = \{S(t)\} \tag{8}$$

where [M] is the mass matrix (it is diagonal with the mass $m_i$ of each DOF i), [C] is the damping matrix, [K] is the stiffness matrix and $\{S(t)\}$ is the applied force.

To solve this linear system, we simultaneously diagonalize matrices [K] and [M] and assume that [C] is also diagonal with the N damping ratios $\xi_j$ on the diagonal. The N eigenvalues $\omega_j^2$, the corresponding eigenvectors $\Phi_j$ and damping ratios $\xi_j$ are the modal parameters of the system. Considering a building forced into vibration by an earthquake, the displacement $V_i(t)$ of each DOF i is the sum of the displacement of the ground $U_S$ and the relative unknown displacement of the structure $U_i$. The solution of Eq. 8 is of the form $\{U(t)\} = [\Phi]\{y(t)\}$.

The equation of motion can be rewritten:

$$y_j''(t) + 2\xi_j\omega_j y_j'(t) + \omega_j^2 y_j(t) = -p_j U_s''(t) \qquad (9)$$

where $\dfrac{\{\Phi_j\}^T[K]\{\Phi_j\}}{\{\Phi_j\}^T[M]\{\Phi_j\}} = \omega_j^2$ are the eigenvalues, $\dfrac{\{\Phi_j\}^T[C]\{\Phi_j\}}{\{\Phi_j\}^T[M]\{\Phi_j\}} = 2\xi_j\omega_j$ and

$\dfrac{\{\Phi_j\}^T[M]\{1\}}{\{\Phi_j\}^T[M]\{\Phi_j\}} = p_j$ is the participation factor of mode j.

The analytical solution of these N independent equations for $y_j(0)=0$ and $y'_j(0)=0$ is known as Duhamel integral, given by:

$$\forall j \in [1,N] \; y_j(t) = \frac{-p_j}{\omega'} \int_0^t U_s''(\tau) e^{-\xi_j\omega_j(t-\tau)} \sin(\omega'(t-\tau)) d\tau \qquad (10)$$

with $\omega_j'^2 = \omega_j^2(1-\xi_j^2)$.

Assuming a constant mass per floor equal to 1000 kg/m$^2$ and knowing the geometry of recording array, it is possible to determine the elastic motion of the system under moderate earthquakes from the knowledge of the modal parameters determined using ambient vibrations, without any assumptions having to be made on the structural design.

In order to evaluate the stiffness matrix from the modal parameters, an assumption on the behaviour of the system must be made. Two models are generally employed for the description of building behaviour [31]: a cantilever beam representing shear-wall RC buildings and a shear beam for RC-frame buildings.

### 3.2 Stiffness matrix form

The shear beam model assumes that the motion at one story depends only on the motion of the stories above and below. The underlying hypothesis is that the floors are much stiffer than the walls. The stiffness matrix [K] can be written as follows:

$$[K] = \begin{vmatrix} k_1 + k_2 & -k_2 & 0 & \cdots & \cdots & 0 \\ -k_2 & k_2 + k_3 & -k_3 & \ddots & & \vdots \\ 0 & -k_3 & \ddots & \ddots & \ddots & \vdots \\ \vdots & \ddots & \ddots & \ddots & \ddots & 0 \\ \vdots & & \ddots & -k_{n-1} & k_{n-1} + k_n & -k_n \\ 0 & \cdots & \cdots & 0 & -k_n & k_n \end{vmatrix} \quad (11)$$

where $k_i$ is the stiffness of the story i.

For the shear beam assumption, the characteristic ratio between the harmonics and the fundamental frequency is [31]:

$$\forall k \in [1,N] \ \frac{f_k}{f_1} = 2k - 1 \quad (12)$$

The equation of the eigenvalues $[K]\{\Phi_i\} = \omega_i^2 [M]\{\Phi_i\}$ for the shear beam model can be then inverted in order to evaluate the stiffness matrix [K].

The corresponding linear system can be solved into the analytical formula:

$$\forall j \in [1,N] \ k_j = \omega_i^2 \frac{\sum_{l=j}^{n} m_l \Phi_{li}}{\Phi_{ji} - \Phi_{(j-1)i}} \quad (13)$$

with $\Phi_{0i} = 0$.

This analytical formula (Eq. 13) allows the extraction of the stiffness values at each floor from the modal shapes. Theoretically, only one mode is necessary to calculate the stiffness matrix under the shear hypothesis. Nevertheless, this formula shows that when two close-together stories have approximately the same deflection for one particular mode, i.e., $\Phi_{ji} \approx \Phi_{(j-1)i}$, the stiffness values are highly sensitive to uncertainties in the modal shape determination. Therefore, considering two modes reduce these uncertainties.

**4. Example of stiffness estimation using experimental data**

To illustrate the use of ambient vibration for the stiffness evaluation of the building, two experiments have been carried out in a dwelling located in the centre of Grenoble (France). The ambient vibrations (AV) and the vibrations induced in this structure by the demolition of a close-by bridge allowed the extraction of the modal parameters of the building using the FDD method. We compared the modal parameters found in both experiments and used them:

- to build a modal model that can be validated by recordings of the bridge's collapse;
- to estimate the stiffness of the building at each story in each direction.

The building is a nine-story reinforced concrete (RC) structure with both frames and shear walls (Fig. 4). It was built in 1939 and it is one of the most important classes of buildings found in the urban area of Grenoble. The concrete is poorly reinforced and it has been built without using a seismic building code. It is located within an urban block, with two walls in contact with the surrounding buildings, which are only separated by filled joints. Following the European Macroseismic Scale (EMS98) [32], it corresponds to the type `Reinforced Concrete Frame Building' which is one of the most vulnerable RC classes. In this study, we used the Cityshark II acquisition system [33], a user-friendly seismological station dedicated to ambient vibrations experiments. It allows the simultaneous recording of 18 channels. Six Lennartz Le3D-5s velocimeters with a flat response in the 0.2-50 Hz frequency band were used.

### 4.1 First experiment: Ambient Vibrations (AV)

For the first experiment, ambient vibrations were recorded for 15 minutes (Fig. 4), at a sample rate of 200 Hz, which is long enough with respect to the Brincker criterion (1000 periods) [27]. The sensors were oriented along the longitudinal direction of the building. Two series of recordings were performed in order to record one point per floor (Fig. 4). The sensors at the top of the building (8$^{th}$ and 9$^{th}$ floors) were kept as references since they have the greatest amplitudes while the others were roving sensors.

Using the FDD method, the two first bending modes in each direction have been extracted (Fig. 5, Tab. 1): at 2.73 Hz and 7.71 Hz in the longitudinal direction and at 2.28 Hz and 8.69 Hz in the transverse direction. Moreover, despite the low level of energy it is possible to identify the frequency of the third bending mode in each direction but not the modal shapes. That may be the same reason for the second transverse mode showing a curious modal shape. For the longitudinal direction, the ratio between harmonics and fundamental frequencies fit the theoretically-assumed sequence 1, 3, 5 for the shear beam model. In the transverse direction, the ratios are 1, 3.8 and 6.8.

This ratio is a characteristic of the model and helps to choose which model provides a closer fit to the building. Our building (Tab. 1) behaves therefore more in bending in the transverse direction than in the longitudinal one. This can also be seen in the experimental modal shapes (Fig. 5) where the first transverse modal shape looks like a cantilever beam. Nonetheless, we used the shear beam model for the two directions of the building, knowing this model represents only roughly the longitudinal direction.

Since the stiffness is proportional to the square of the frequency, the first frequencies tell us that the longitudinal direction is 40% stiffer than the transverse one.

The FDD method shows other modes that may correspond to torsional behaviour Tab. 1. The irregular position of the shear walls and beams induces undoubtedly this kind of torsional motion, that could be confirmed using several recording points at the same floor.

## 4.2 Second experiment: Bridge Demolition (BD)

In order to check the relevancy of the ambient vibrations for the evaluation of the modal parameters, the second experiment was focused on building motion evaluation for ground excitation. This motion was induced by the controlled demolition of a bridge located 40 m away from the building (Fig. 6). The same instrument was employed for this second experiment. Because only six Lennartz Le3D-5s sensors could be plugged into the CityShark II, we installed them in six of the nine stories of the building (Fig. 9). 200 s of signal were kept and the sampling rate chosen for these recordings was 100 Hz.

The instantaneous collapse of the deck (source time duration of about 5 ms) generated a vertical seismic motion with a frequency peak around 11 Hz (Fig. 8). The horizontal peak ground acceleration (PGA) recorded in the ground floor of the building was 0.025 m/s$^2$ (Tab. 2). This value is greater than the PGA recorded during recent Alpine earthquakes by the French Accelerometric Network (http://www-rap.obs.ujf-grenoble.fr) in the Grenoble basin. Nevertheless, it is sixty times lower than the design acceleration values enforced in the present code [33] for the Grenoble area (1.5 m/s$^2$). The maximum horizontal acceleration at the top of the building was 0.061 m/s$^2$.

Although the basic assumption of white noise is required, the FDD method was also used to determine the modes of the structure for these explosion recordings. The following results show that neglecting the white noise assumption does not greatly affect the modal parameter determination, as already showed by Ventura [35]. Only the first bending mode in each direction (Fig. 9) and the first supposed torsion mode can be found. The transverse mode exhibits a non-negligible torsion part in its modal shape that was not seen in the AV recordings. This explains the low correlation observed between the two experiments (Fig. 9). The first bending and torsion modes closely match for the two recordings with a difference of less than 1% in frequency (Tab. 3). The MAC values show a very high similarity between the mode shapes obtained in the two experiments (Tab. 3). Some points of the modal shape were linearly interpolated for the bridge demolition because only six points could be simultaneously recorded.

The modal parameters extracted from ambient vibrations and the induced ground motion match so well that we can conclude on one hand that the building responded totally elastically to the ground motion and on the other hand that both modal identifications were reliable, even if in the second case the assumption of white noise input is not valid and the length of the signal limited.

**4.3 Simulation of the building's response to earthquakes and the calculation of stiffness**

In order to validate the aforementioned methodology, we first verified if the motion of the building when forced into vibration by the bridge demolition would be well predicted by AV modal modelling. Once the experimental mode shapes are defined, and by following the section 3.1, the building motion and its behaviour under earthquakes are simulated using Eq. 10 and the experimental modal parameters extracted from the ambient vibrations. In this study, only the two first bending modes in each direction were used. We estimated the corresponding damping ratio using the random decrement method [36] as 4% for all modes except for the first longitudinal mode (where 2% is found).

The displacements computed and recorded at the top of the building are displayed on Figure 10. We observe a good fit between the data and the modelling. Even if the complexity of the building behaviour is not integrated in the model, the simulation reproduces the level of amplitudes as well as the duration, the frequency and the phase of the building motion. It is also possible to compute the inter-story drift by difference between the time history displacements at consecutive floors. The maximum drift Fig. 11 has been calculated in this way using the modal model and the recordings for the bridge demolition. An average of the model in non-recorded stories has been performed to compare the same quantities. The model reproduces the magnitude of the maximum drift but in the longitudinal direction, the drift at the upper stories and especially the fourth is under-estimated.

Nevertheless, these comparisons confirm the relevancy of the modal parameters of buildings extracted from ambient vibrations. Hence, once the modal parameters are known, it is possible to evaluate the ability of the building to resist (or not) the seismic loading corresponding to any a priori ground motion scenario. If we suppose that the maximum drift sustained by concrete shear walls controlled by shear without any damage is $4 \times 10^{-3}$ [37], we are able to determine for a chosen ground motion if the building reaches this threshold or not. As expected, in the case of the bridge demolition experiment Fig. 11, the maximum drift remains very low (around $6 \times 10^{-6}$). It is quite similar in the longitudinal and transverse directions. The model outlines that the third floor is subjected to a greater drift in the transverse direction but does not show similar behaviour in the fourth story in the longitudinal direction.

Assuming a shear model for this building, as suggested by the ratio $\frac{f_2}{f_1} \approx 3$, we can deduce the stiffness at each floor following the Eq. 13. The results are displayed on Fig. 12. The stiffness at each story is between 20 and 50% larger for the longitudinal direction than for the transverse one. In the longitudinal direction, because of the quality of the modes extracted from the ambient vibrations, two modes were available to determine the stiffness. In this direction, the only difference between the 2 modes appears for the third story. As mentioned in section 3.2, this difference reflects the sensitivity to the error at the anti-node of the second mode, when the deflection is lower. For the last floor, considering the first mode shows the same imprecision in the stiffness evaluation. This illustrates the necessity to simultaneously consider more than one mode in order to increase the quality of the stiffness evaluation. It shows also the necessity in estimating the uncertainty of the extracted modal shapes. It should be noticed that even if the amplitudes vary slightly, these three independent stiffness shapes show the same variations along all the stories. This assures that these stiffnesses are valid, for example the third story is softer than the second and the fourth. This can be the reason for the greater strain recorded (and modelled) for this story (Fig. 11).

From these results, a linear trend of stiffness, decreasing with height, in both directions is evident for this building. It is not a very irregular configuration regarding seismic vulnerability. However, the third story has been identified as a "soft" story. In case of an earthquake, the building must resist a larger drift at this story making it more vulnerable than other buildings of the same RC class.

## 5. Conclusions

Herein we demonstrated the utility of a set of simple techniques using ambient vibration recordings to complement seismic vulnerability assessment investigations of existing buildings. Decomposing the motion of a building into simple modes (bending and torsion) is the first step in the assessment of its behaviour under an earthquake. We computed, therefore, the modal parameters using the FDD method, which is easy to perform and reliable even in case of close modes. The experiments showed that the FDD is able to extract reliable information (frequency, modal shapes and eventually damping) even if the basic assumption of white noise is not fulfilled. The values of the measured frequencies are directly linked to the stiffness of the building so that we can assess the average stiffness ratio between the longitudinal and the transverse directions. For vulnerability assessment this is not sufficient so we used a simple lumped-mass model with an assumption on the stiffness matrix (shear beam) in order to calculate the stiffness at each story. The choice of this stiffness model is based on the sequence of the resonance frequencies of the studied building. The inversion of the stiffness matrix is analytical in this case and two modes are practically needed for a reliable stiffness determination. Having the stiffness at each story, we can identify soft stories or bad structural configurations, which increase the vulnerability of the building to earthquakes. We showed, for example, that the studied building displayed an irregularity in the third story, which makes it more vulnerable to earthquakes than other buildings of the same type. The presented method is therefore able to identify the most vulnerable buildings within a building class. The modal model, free of any assumptions on the stiffness behaviour of the building, allows the computation of its elastic motion subjected to weak to moderate earthquakes and then the inter-story drift. Using directly this drift or by applying the integrity threshold concept developed by Boutin et al. [9] for reinforced concrete, one can decide whether the building reaches the inelastic range, i.e. the beginning of permanent damage. Several earthquake scenarios can also be applied in order to identify for what acceleration levels the building stays within its elastic domain.

**List of tables**

Table 1. Frequencies of the building modes extracted from AV using the FDD technique.

| Mode | Frequency (Hz) | $f_k/f_1$ |
|---|---|---|
| Transverse 1 | 2.28 | 1 |
| Longitudinal 1 | 2.73 | 1 |
| Torsion 1 | 4.74 | |
| Torsion 2 | 5.64 | |
| Transverse 2 | 8.69 | 3.8 |
| Longitudinal 3 | 12-13 | 4.5 |
| Transverse 3 | 15.5 | 6.8 |

Table 2. Maximum accelerations recorded for the bridge demolition (m/s$^2$).

| Direction | Ground Floor | Top floor |
|---|---|---|
| Longitudinal | 0.025 | 0.051 |
| Transverse | 0.019 | 0.062 |
| Vertical | 0.046 | 0.094 |

Table 3. Comparison of frequencies and modal shapes (MAC value) extracted from Ambient Vibrations (AV) and Bridge Demolition (BD) experiments.

| Mode | AV | BD | Comparison | |
|---|---|---|---|---|
| | f (Hz) | f (Hz) | Freq. Var. | MAC value |
| Transverse 1 | 2.28 | 2.27 | -0.4% | 85% |
| Longitudinal 1 | 2.73 | 2.75 | 0.7% | 99% |
| Torsion 1 | 4.74 | 4.77 | 0.6% | 97% |
| Transverse 2 | 8.69 | 8.46 | -2.6% | 20% |

**List of figures**

Figure 1. Overview of this study's scope. We relate, in a simple way, experimental modal parameters of a building (frequencies, modal shapes and damping) to ground-motion parameters (earthquake scenario) and structural parameters (stiffness) for seismic vulnerability assessment

Figure 2. Practical use of the Frequency Domain Decomposition (FDD) method on synthetic data of a Gaussian random signal convolved with a sine function of frequency 0.1 (simulation of a 2-story building) (a). The singular value decomposition (c) extracts redundant information in the Power Spectral Density (PSD) matrices (b). The obtained modal shape is exactly the preassigned one (d).

Figure 3. Validity range of the building mode extracted by FDD method using a MAC value greater than 80%. The solid line is the normalised first singular value of PSD matrices of ambient vibration recordings in a building. The bold line (first singular vectors with a MAC value greater than 80% except for two unused points) is one of the estimated mode of the structure. The crosses display the MAC values associated with this mode.

Figure 4. Left: The studied building ; Centre: The experimental scheme for the first and second Ambient Vibration (AV) datasets ; Right: Time history acceleration deduced from AV velocity recorded at the eighth floor.

Figure 5. Building frequencies selected in the first singular values of the power spectral density matrices computed from AV recordings (left) and corresponding modal shapes (right)

extracted by the FDD method (bending modes only). For each identified mode, the bold bell corresponds to a MAC value greater than 80%.

Figure 6. Aerial location of the bridge demolished in July 2004 (A) and the study-building (B).

Figure 7. Accelerations produced by the bridge demolition in the structure: time-histories in Longitudinal (left) and Transverse (right) directions at the ground and top floors.

Figure 8. Fourier amplitude spectra of the recordings of the bridge demolition at the ground floor of the structure. The spectra are smoothed using the function described in [37]. The modal frequencies of the structure deduced from ambient vibrations are labelled X1 and Y1.

Figure 9. Comparison of the first modal shapes extracted from AV and BD (left) and experimental scheme during the BD (right). The stories that were not measured are linearly interpolated.

Figure 10. Comparison of time histories (left) and spectra (right) at roof level between the observations and the modelling of the building response to the bridge demolition.

Figure 11. Comparison of the drift envelope at each available story modelled and recorded in the structure during the bridge demolition. Solid and dashed lines are the longitudinal and transverse motion, respectively. Thin and bold lines are for recordings and modelling, respectively.

Figure 12. Stiffness of the building calculated from the AV modal shapes. The second transverse mode gives non-physical results.

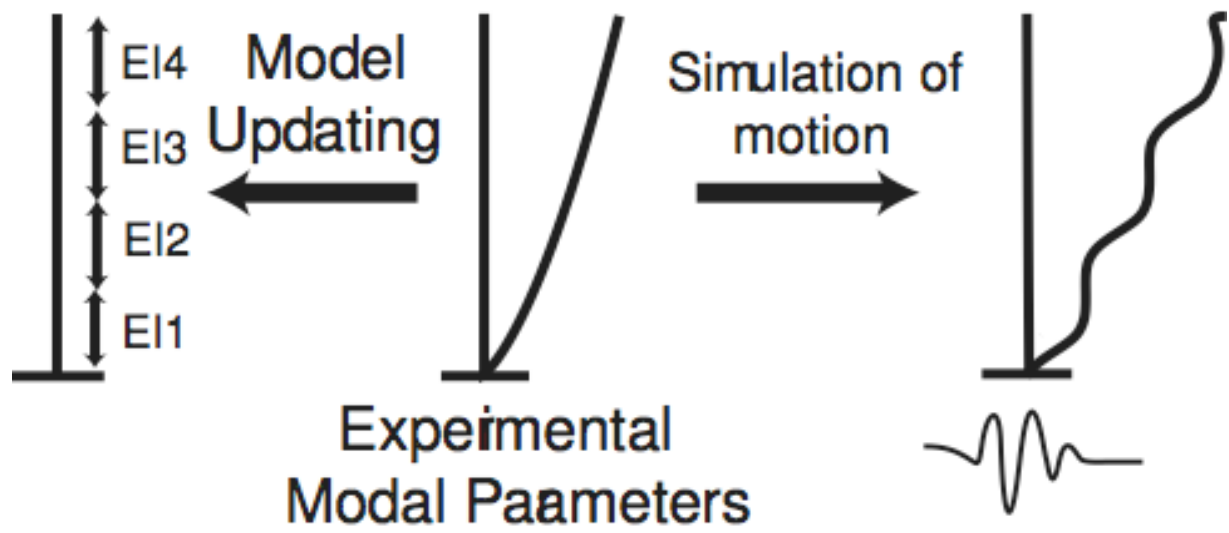

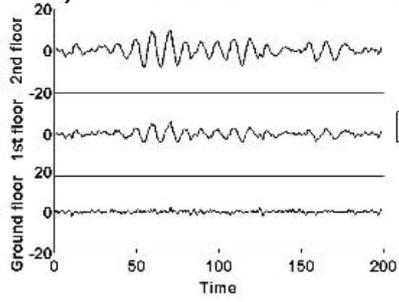
a) Simultaneous recordings

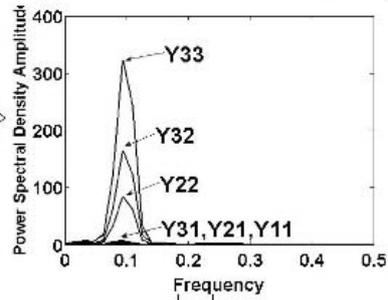
b) PSD Matrices Y (3x3)

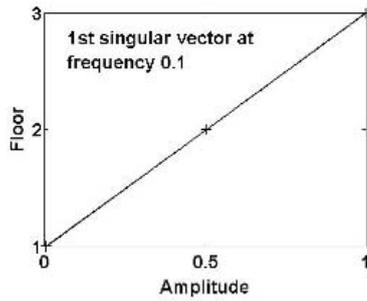
d) Modal parameters

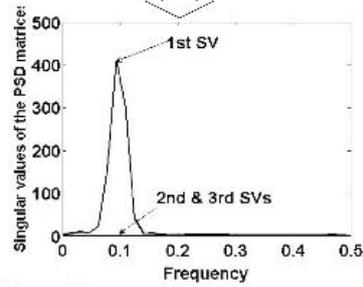
c) Singular value decomposition

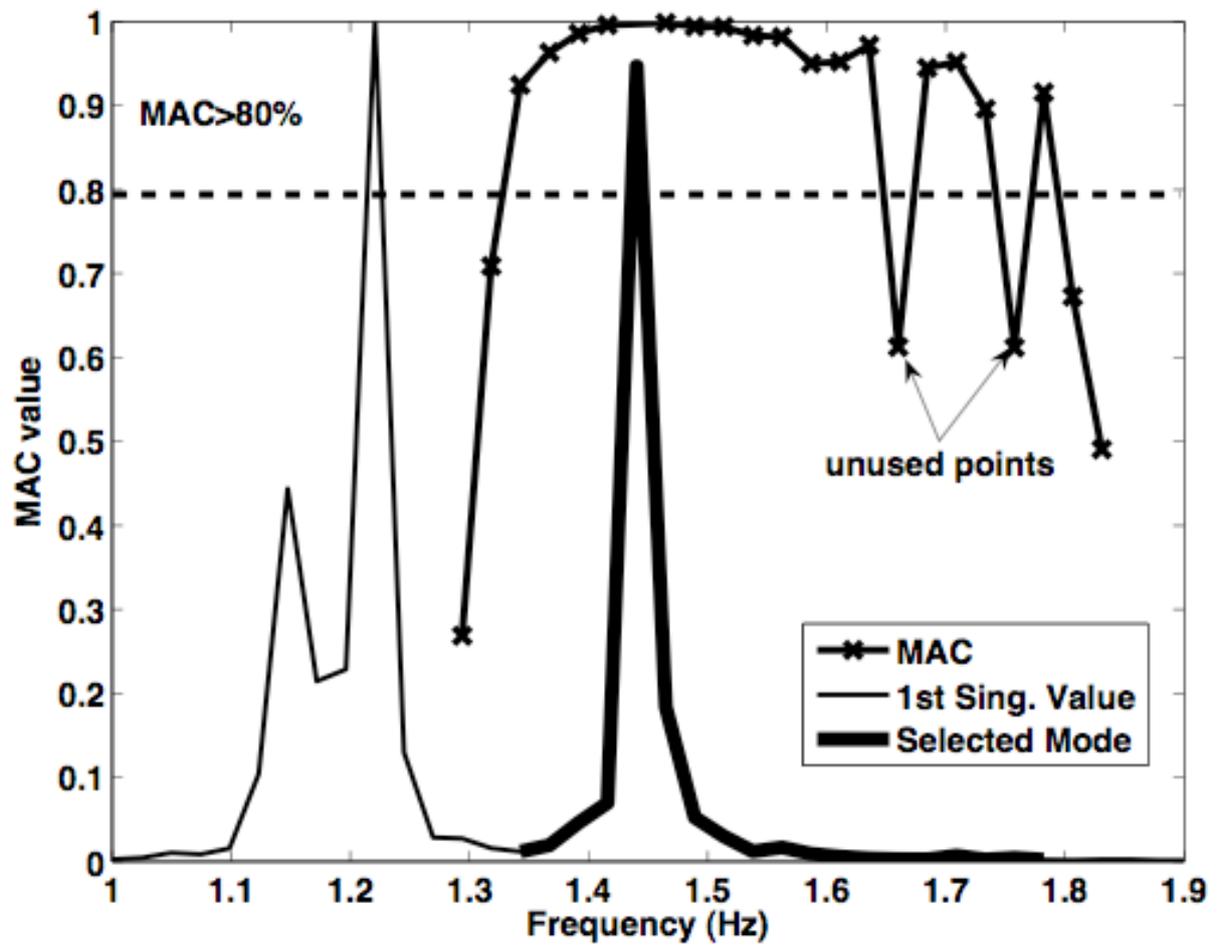

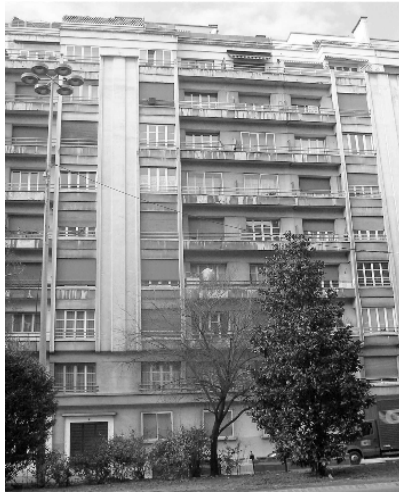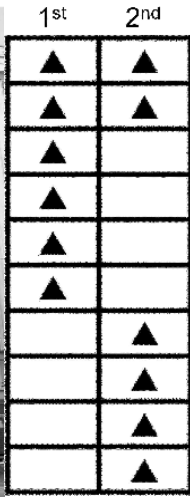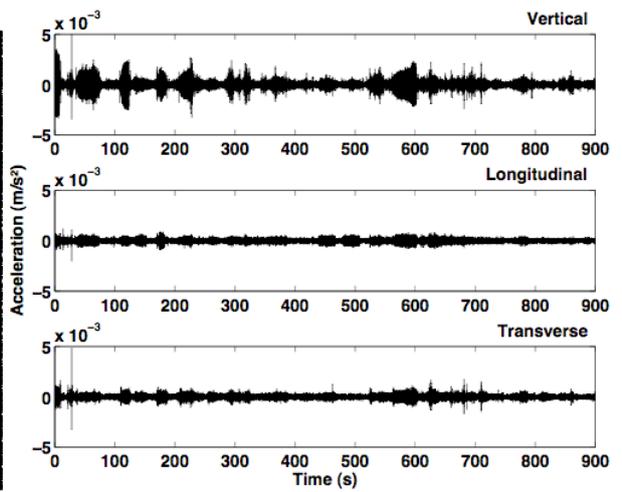

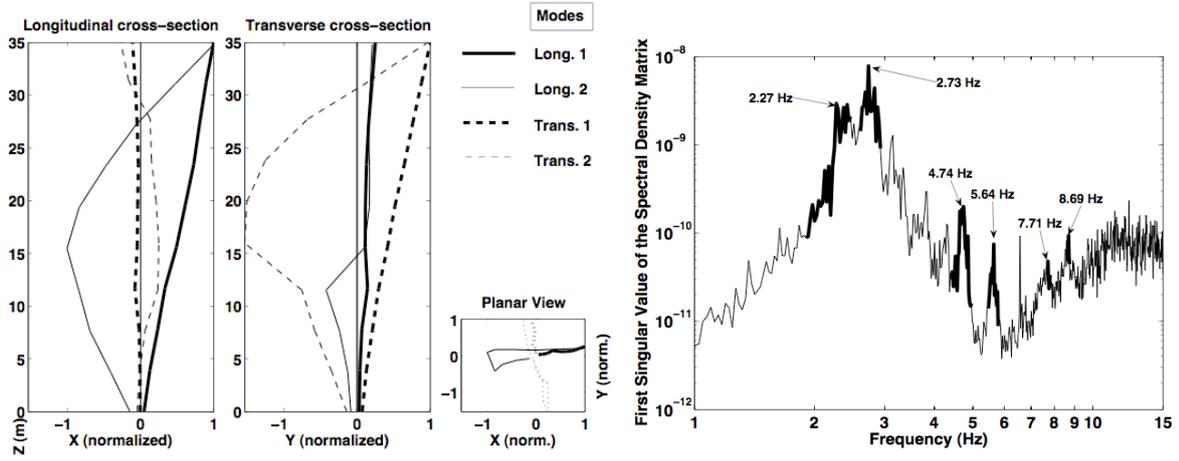

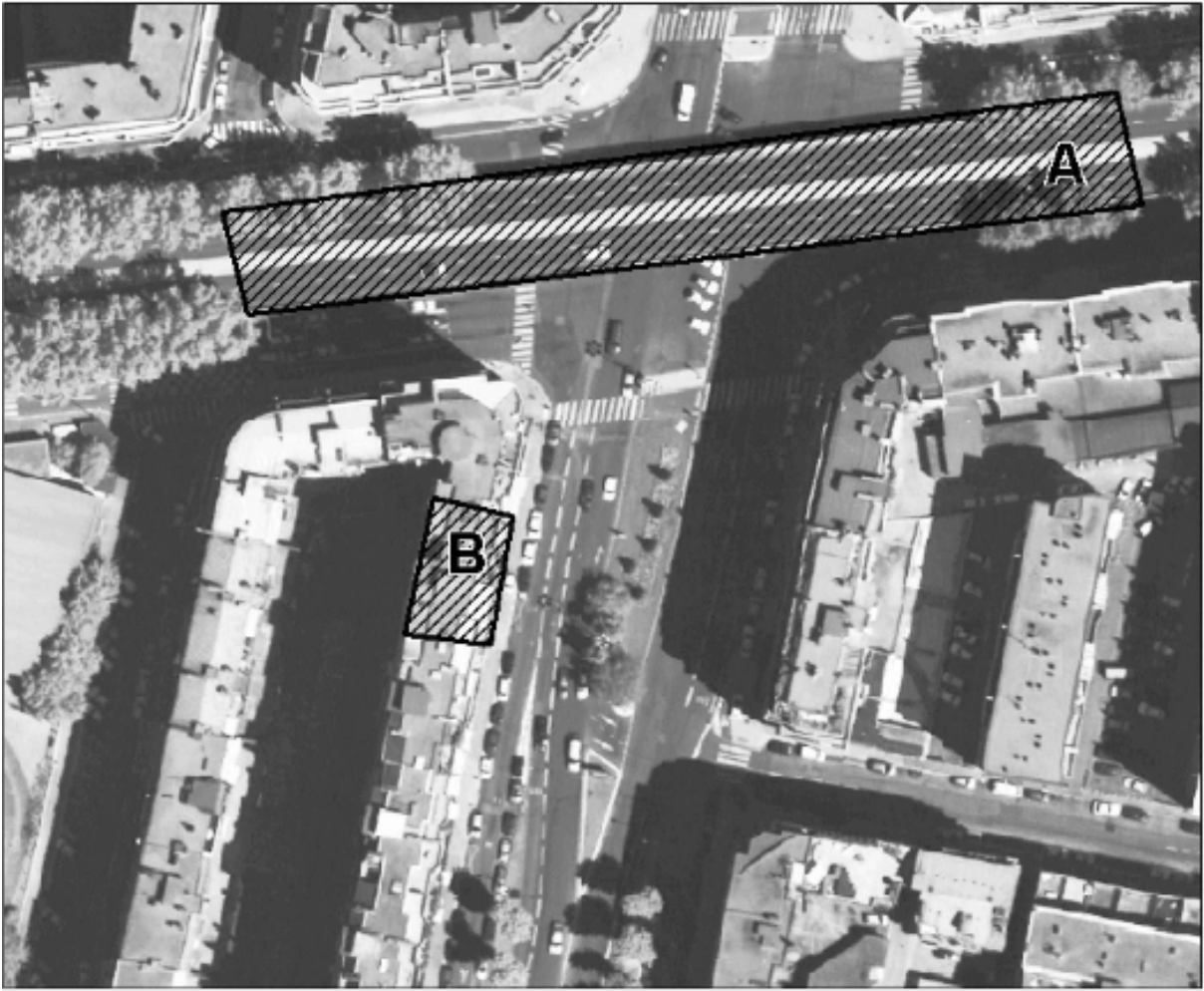

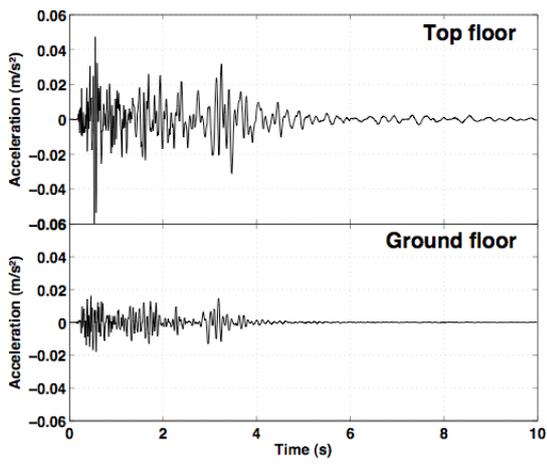 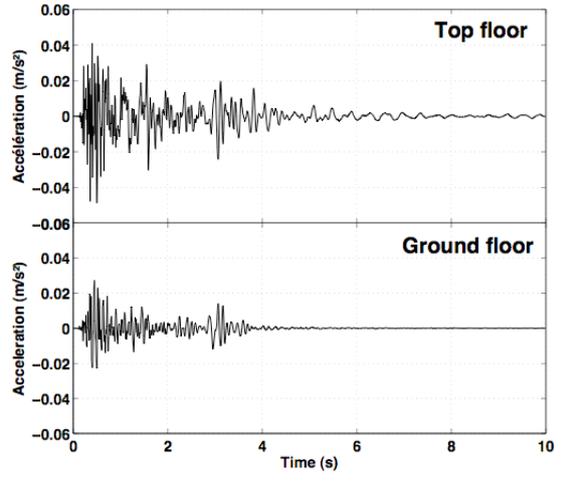

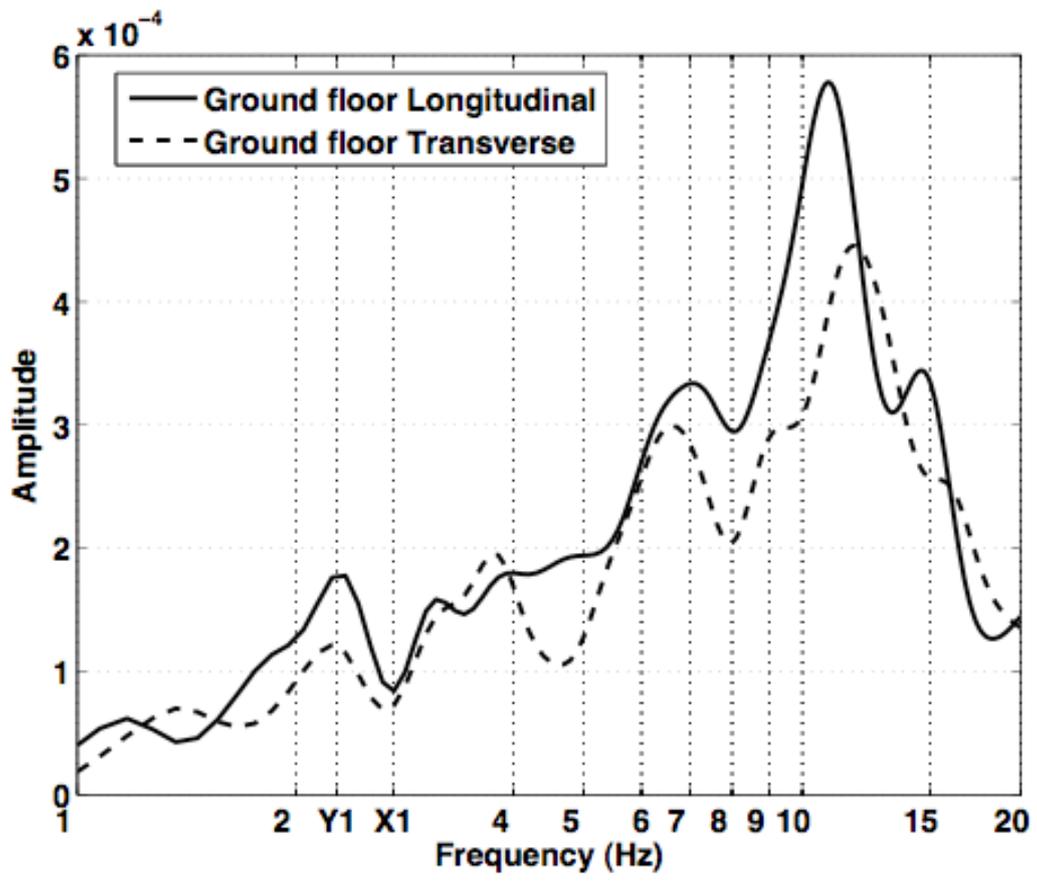

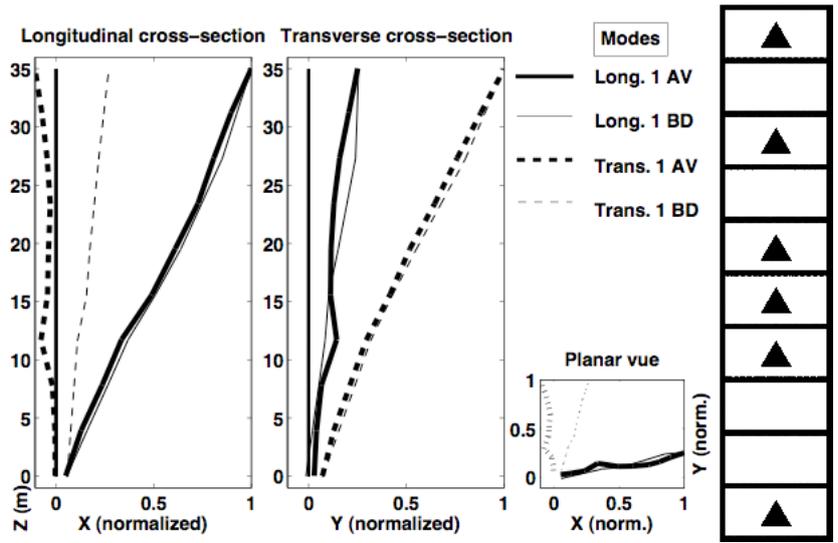

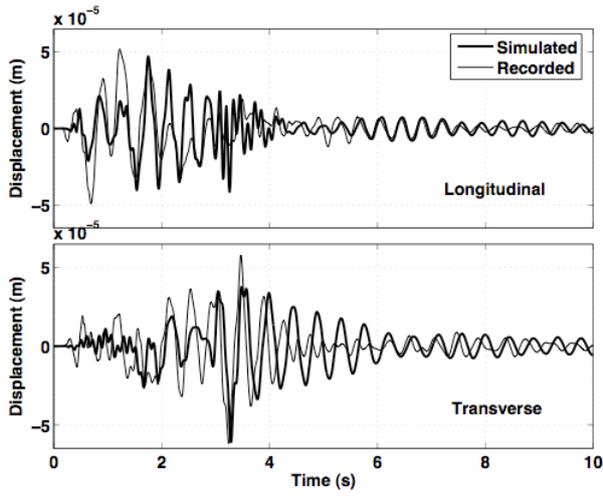
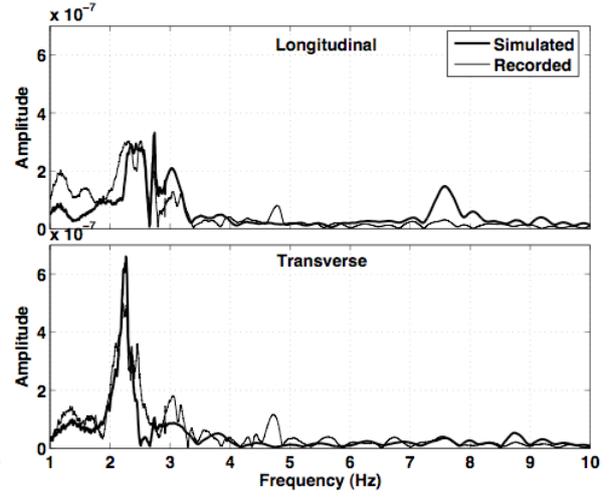

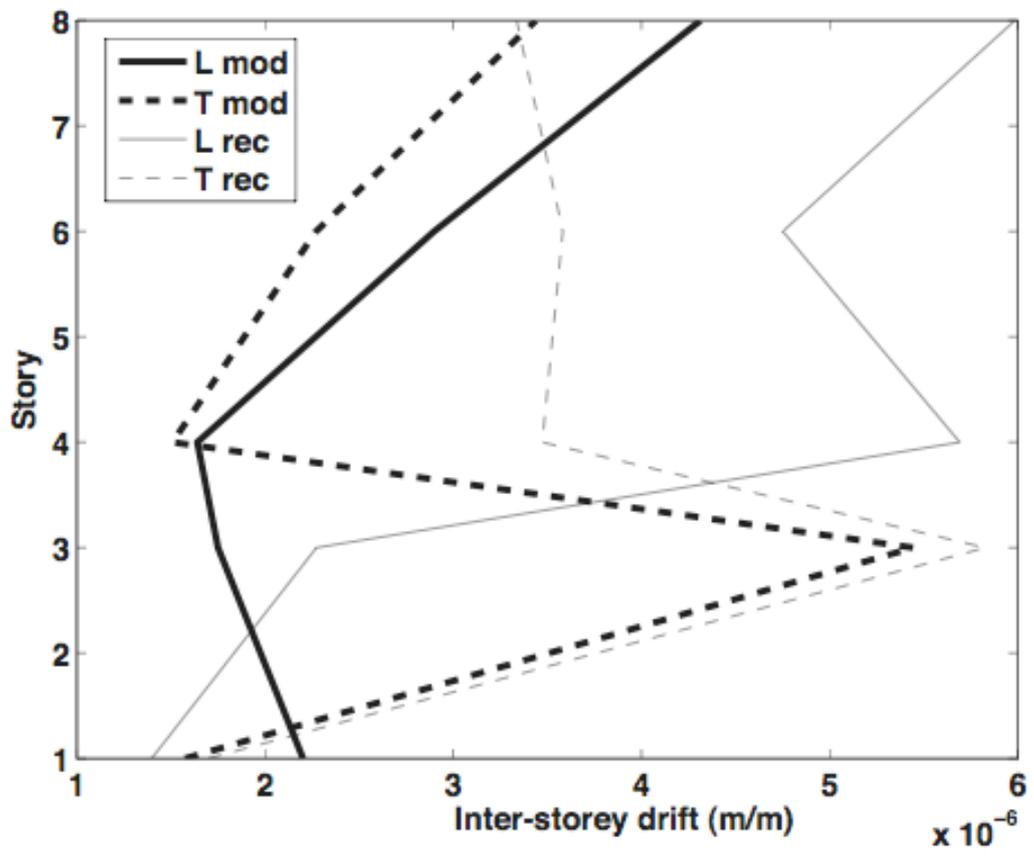

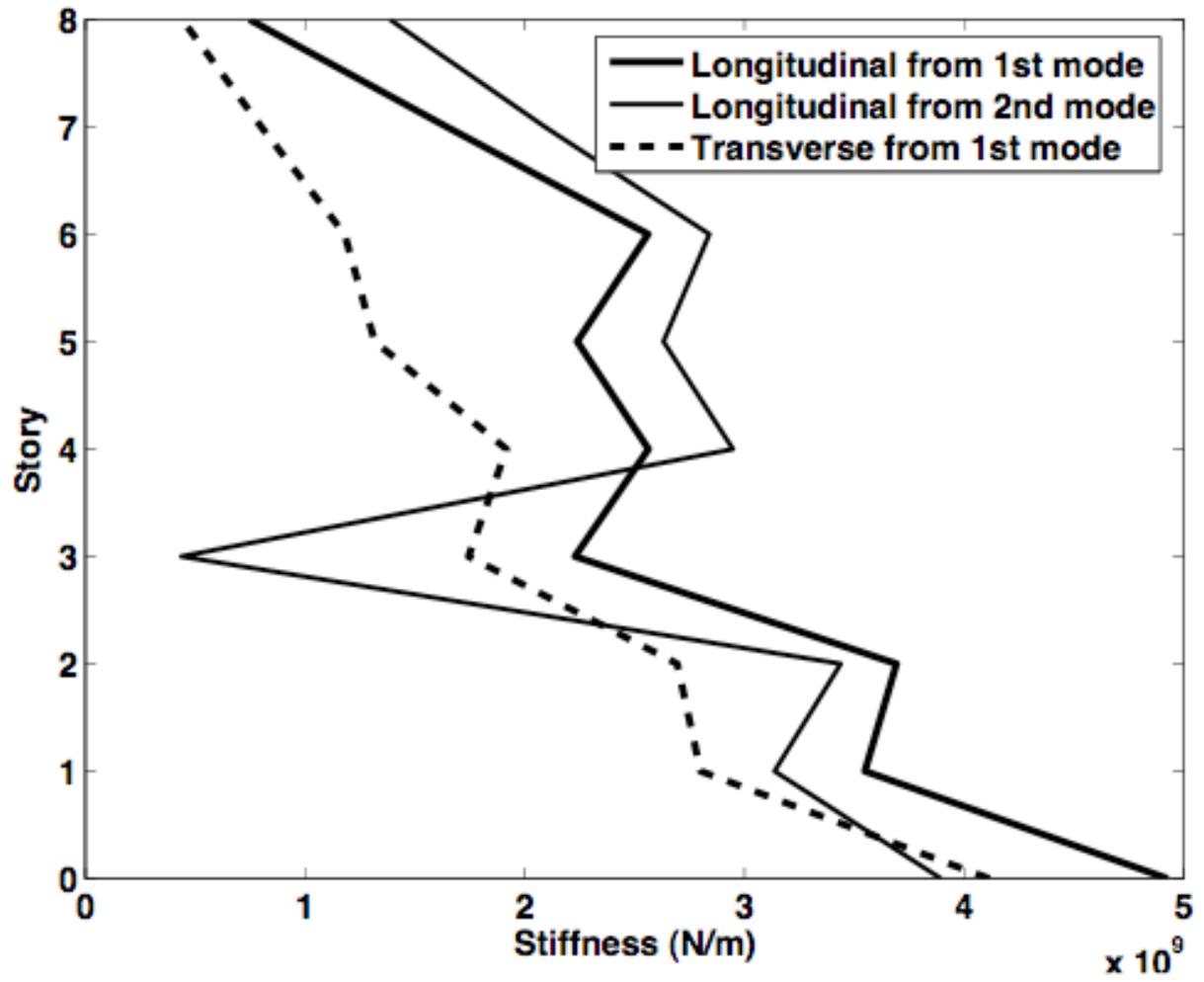